\documentclass[preprint]{aastex6}

\shorttitle{Zodiacal light with Pioneer 10}
\shortauthors{Matsumoto et al.}

\begin{document}

\title{Zodiacal light beyond Earth orbit observed with Pioneer 10}

\author{T. Matsumoto\altaffilmark{1,4}} \author{K. Tsumura\altaffilmark{2}} \author{Y. Matsuoka\altaffilmark{3}} and \author{ J. Pyo\altaffilmark{4}}
\

\affil{\altaffilmark{1}Department of Space Astronomy and Astrophysics, Institute of Space and 
Astronoutical Science, Japan Aerospace Exploration Agency, Sagamihara, Kanagawa 252-5210, Japan}
\affil{\altaffilmark{2}Frontier Research Institute for Interdisciplinary Science,
Tohoku University, Sendai, Miyagi 980-8578, Japan}
\affil{\altaffilmark{3}Research Center for Space and Cosmic Evolution, Ehime University, Matsuyama, Ehime 790-8577, Japan} 
\affil{\altaffilmark{4}Korea Astronomy and Space Science Institute (KASI), Daejeon 34055, Korea}

\begin{abstract}

We reanalyze the Imaging Photopolarimeter (IPP) data from Pioneer 10 
to study the  zodiacal light in the {\it B} and {\it R} bands beyond Earth orbit,
applying an improved method to subtract integrated star light (ISL) and diffuse Galactic light (DGL).
We found that there exists a significant instrumental offset, making it difficult to examine the absolute sky 
brightness. Instead, we analyzed the differential brightness, i.e., the difference in sky brightness from the
average at high ecliptic latitude, and compared with that expected from the model zodiacal light.
At a heliocentric distance of $r< 2$ au, we found a fairly good correlation between the {\it J}-band 
model zodiacal light  and the residual sky brightness after subtracting the ISL and DGL.
The reflectances of the interplanetary dust (IPD) derived from the correlation study are marginally 
consistent with previous works. 

The zodiacal light is not significantly detectable at $r>3$ au, as previously reported. However,
a clear discrepancy from the model is found at $r=2.94$ au which indicates the existence of a local dust cloud
produced by the collision of asteroids or dust trail from active asteroids (or main-belt comets).

Our result confirms that the main component of the zodiacal light (smooth cloud) is consistent with the model 
even beyond the earth orbit, which justifies the detection of the extragalactic background light (EBL) after subtracting
the zodiacal light based on the model.

\end{abstract}

\keywords{zodiacal dust - interplanetary medium - minor planets, asteroids: general}

\section{Introduction}  

Pioneer 10 is a NASA deep-space probe launched on 1971. 
After encountering  Jupiter, it left for interstellar space. One scientific instrument on board Pioneer 10 
is the Imaging Photopolarimeter (IPP; \citet{Pellicori73}) which was designed to observe the zodiacal light 
during the cruising phase and has two filter bands, the {\it B} band 
(blue: 395-485 nm) and the {\it R} band (red: 590-690 nm) each with a  $2.3 \times 2.3$ square degree field of view.
Pioneer 10 is a spinning spacecraft, and the IPP scanned the sky by offsetting the telescope axis 
from the spin axis. 

Figure 1 shows the coordinate system used in this paper.
The distance to the observer (Pioneer 10) from the sun is defined as heliocentric distance, $r$. 

The IPP data have been studied by many researchers to examine the zodiacal light and 
interplanetary dust (IPD). \citet{Hanner74} obtained the dependence of sky brightness 
on $r$ at the specific fields on the ecliptic plane, reporting that
the zodiacal light extends to $r\sim3$ au.  \citet{Hanner76} 
studied the radial distribution of the interplanetary dust and found that a model with power law, $r^{-1}$,
 describes the observations well.

\citet{Matsuoka11} reanalyzed the IPP data based on recent progress in deep star 
counts and measurement of the diffuse Galactic light (DGL). They, however, 
concentrated on the detection of extragalactic background light (EBL) and used the IPP data beyond the main belt.
Following their method, we analyze all IPP data  to obtain the radial distribution of the zodiacal light from 1 to 5 au.
We compare them with the widely used model of the zodiacal light based on the DIRBE observation \citep{Kelsall98} 
and examine differences from the model.
 
This paper is organized as follows. In \S2, we provide information on the observation and describe the 
data reduction procedure. \S3 shows the radial distribution of the zodiacal light and its correlation
with that of the model.  In \S4, we indicate the implications of our result for the study of the
zodiacal light and the EBL. Finally, \S5 summarizes our results.  

\section{Validation of the IPP data}

The quality of the IPP data is rather low due to the confusion of stars, and the low signal-to-noise ratio.
 Recent progress on deep star counts and measurement of DGL have enabled us to perform an improved 
 analysis of the IPP data. 
 
 Following \citet{Matsuoka11}, we first remove the extraordinary data from the IPP data set and 
 retrieve the data at high Galactic 
 latitude ( $|b| > 35^{\circ}$) to avoid the bright Galactic plane.  We then subtract the integrated star light (ISL) and
DGL. Since details are shown in \citet{Matsuoka11}, we describe the
 procedure only briefly.
 
 In the original data set, the contributions of stars brighter than $V=8$ mag are already subtracted. 
The ISL due to relatively bright  stars ($V<11$ mag) is estimated using the Tycho-2 Catalog \citep{Hog00},
while those of stars fainter than $V=11$ mag are derived from the HST Guide Star Catalog II version 2.3 (GSC-II;  \citet{Lasker08}). 
We calculate the contribution of stars even 
 fainter than the detection limits of GSC-II using a star count model provided by the stellar population synthesis code
 TRILEGAL \citep{Girardi05}. We estimate the DGL contribution by separating out the component correlated 
 with diffuse Galactic far-infrared emission. The applied conversion factors from the far-infrared to the 
 optical bands are consistent
 with recent DGL observations by AKARI  \citep{Tsumura13a}, CIBER  \citep{Arai15} and HST \citep{Kawara17}. 
 We then subtract the ISL and DGL from the observed sky brightness to obtain the residual brightness,
  which consists of the zodiacal light and isotropic emission.  
  
Zodiacal light is a sunlight scattered by IPD particles, and has been studied by many researchers
( \citet{Leinert98}, \citet{Levasseur-Regourd01}, \citet{Buffington16}). DIRBE on board
COBE observed the zodiacal light extensively from near-infrared to far-infrared wavelength 
regions \citep{Hauser98}. Based on the DIRBE data, \citet{Kelsall98} constructed the
physical model of the IPD using seasonal variations of the zodiacal light that are caused by the difference of the
symmetrical surface between the earth orbit and the zodiacal dust cloud. 
To compare the IPP data, we choose the model brightness at the shortest wavelength band, the {\it J} band 
 (1.25 $\mu m$, $\Delta\lambda=0.31\mu m$), because 
the thermal emission of the IPD is negligible there.
 
We calculate the brightness of the {\it J}-band model zodiacal light for all observed fields at the same epochs based on the model by \citet{Kelsall98}. 
Since the model provides the zodiacal light observed from Earth, we modify it to calculate 
the zodiacal light from anywhere in the solar system.
We do not change the spatial distribution of the IPD in the solar system
and apply the same cut-off radius of $r=5.2$ au in the model for integration along the line of sight.

Figure 2 shows a typical correlation diagram of the {\it R}-band residual brightness with the {\it J}-band
model brightness at $r=2.41$ au.
We present the surface brightness as $\lambda \cdot I_{\lambda}$ in unit of $nWm^{-2}sr^{-1}$ throughout the paper.
Since clearly abnormal data are present, we clip the data points exceeding $3\sigma$ over the best-fit line. 
A similar clipping procedure is applied for the subgroups (ecliptic plane, high-ecliptic latitude region, and 
red-to-blue ratio, etc.). This clipping procedure reduces the number of data points, however, does not 
affect correlation studies with the model.   

Figure 3 indicates the average residual sky brightness at high ecliptic latitude ($|\beta|> 40^{\circ}$) 
and toward the anti solar direction ($\varepsilon>90^{\circ}$, where $\varepsilon$ is the solar elongation angle). 
This sky area is chosen to obtain enough number of data points for a good statistics with a small
dispersion.
Figure 3 clearly shows the existence of negative values at large heliocentric distances which are unrealistic.
Since sky brightness at such distances is fairly isotropic, we regard these negative values as being
caused by instrumental offsets, which depend on the observed epoch. 
Furthermore, we find a significant correlation between the {\it B}- and {\it R}- band noises 
suggesting that the IPP suffered from extra noise due to external signals.

Due to the instrumental offset, we cannot regard the residual brightness as  absolute sky brightness,
and have to use only the relative variation of the residual brightness at each epoch for the study of the zodiacal light.

Another problem in the IPP data is a stray light. At $r= $2.29, 2.41, and 2.46 au, 
bright excess emission is found on the ecliptic plane at small solar elongation angles. Since this excess 
emission appears at the specific angle between the sun and the spin axis ($21^{\circ} \sim 25^{\circ}$)  and 
no physical origin is known for this excess emission, we regard it as stray light. The IPP data of  $\varepsilon<90^{\circ}$
at these heliocentric distances are not used in the following analysis.

\section{Radial dependence of the zodiacal light }

To avoid effects of instrumental offset, we define the differential brightness, that is, 
the difference of the sky brightness  from the average brightness at high ecliptic latitude ($\varepsilon>90^{\circ}$, $|\beta|> 40^{\circ}$, Figure 3).
The differential brightness is independent of the offset problem, since instrument offset is stable in 
each observation period.  On the other hand, the contribution of isotropic emission disappears in the differential brightness.
Hereafter, we perform our analysis based on this differential brightness for all the observed fields both for the IPP data  and 
the {\it J} band model brightness. 

Figure 4 shows the radial dependence of the average differential brightness on the ecliptic plane 
($120^{\circ}<\varepsilon<150^{\circ}$, $|\beta|<10^{\circ}$).
We concentrate in the outer solar system but avoid the effect of
Gegenschein, since it was not observed by DIRBE. In Figure 4, we did not include the results at $r=1.14$ and $1.17$ au, 
since the data number in these epochs are too small to
obtain reliable results. The position of the main belt (the asteroid belt) is indicated by the thick green line.

According to Figure 4, the zodiacal light declines very smoothly toward the outside of the solar system. 
Figure 5 indicates the correlation of the same data set with the {\it J}-band model brightness; an excellent
correlation is found. The result of the linear fit is indicated by the solid red line and the dotted blue line for the {\it R} and {\it {\it B}} bands, 
respectively. The ratios of the {\it B}- and {\it R}-band brightness to the {\it J}-band model brightness are $1.25 \pm 0.06$, $1.94 \pm 0.1$, respectively,
while solar values are 1.39 and 1.82, respectively.
Good correlation in Figure 5 implies that the spatial distribution and optical properties of the
IPD are well-represented in the model \citep{Kelsall98}.
The relative reflectances to the {\it J} band are shown together with previous work ( \citet{Matsumoto96},  \citet{Tsumura10},  
\citet{Kawara17}) in Figure 6. 
The albedo (ratio of the scattered flux to the input flux at the specific wavelength) is also shown in right ordinate based on the {\it J}-band albedo of the model \citep{Kelsall98}, assuming scattering phase
functions are same in these wavelength range.
Considering the large errors of the IPP data, obtained reflectances are marginally consistent
with previous works.  

The features of the zodiacal light at large heliocentric distances are not so clear.  Figure 4 indicates the 
zodiacal light  is not detected at  $r>3$ au, which is consistent with the previous report \citep{Hanner74}. 
However, this does not imply the inconsistency with the model because of small signal-to-noise rations. 
On the other hand, we found a clear discrepancy from the model toward the tangential directions at $r=2.94$. 
Figure 7 shows the zodiacal light at 
$|\beta|<10^{\circ}$  is depended on the solar elongation angle, $\varepsilon$, at $r=$2.64 and 2.94 au 
both for the leading (right) and trailing (left) directions.  
The central part of the Figure 7 shows the trajectory of Pioneer 10 on the ecliptic plane and the range of the line of sights 
of the IPP at both epochs.
As for the leading directions at both epochs, it is difficult
to obtain a definite result due to the large errors.
The zodiacal light at $r=2.64$ au towards the trailing direction (top left) shows a similar feature as that at $r<2$ au
in Figure 5. On the other had, the zodiacal light at $r=2.94$ toward the trailing direction (bottom left) shows a quite different feature. 
The brightness at the {\it B} band is almost same as that at the {\it R} band, and both bands show
fainter brightness than those estimated from the {\it J}-band model brightness assuming the colors at $r<2$ au. 
 This new feature is characterized with the color, a ratio of
the brightness at the {\it B} band to that at the {\it R} band. This ratio towards the trailing direction at $r=2.94$ au is $1.0 \pm 0.15$,
which is significantly bluer than that of the solar color, 1.3, and that of the inner zodiacal light, $1.55 \pm 0.1$.

Figure 8 shows the correlation of the {\it B}- and {\it R}-band differential brightness with the {\it J}-band model brightness at $r=2.94$ au, in which 
the solid and dotted lines are the same as ones in Figure 5. 
The differential brightness of the zodiacal light at $r=2.94$ au is fainter than 
that expected from the color at $r < 2 $ au.
However, this does not directly imply that the dust density at $r=2.94$ au is lower than that in the model, since the
plotted brightness is a differential brightness; i.e., a difference of the brightness between high $\beta$ and low $\beta$.
It is probable that the zodiacal light at $r=2.94$ au is more isotropic than in the model; in other words, 
the dust layer at $r=2.94$ au is thicker than that in the model. 

\section{Discussion}
A peculiar feature of the dust cloud observed at $r=2.94$ au is important to understand the nature and origin of 
the IPD. The observed evidences indicate that the Pioneer 10 encountered a local dust cloud at $r=2.94$ au.
A filled black circle in the top middle panel in Figure 7 indicates the position of this dust cloud when the IPP observation
was made at $r=2.64$ au, assuming the Kepler motion. The dust cloud is outside of the observed range, which
ensures that the dust cloud was not detected at $r=2.64$ au. 
The extent of the dust cloud is not clear, because the detections of the zodiacal light beyond $r=2.94$ au and those toward 
the leading directions at $r=$2.64 and 2.94 au are not so significant.

There are some reports on the observation of the similar dust cloud.  
\citet{Dermott84} and  \citet{Nesvorny03} explained the dust band discovered by IRAS \citep{Low84} with the thermal emission of the dust 
produced by the collision of asteroids. 
\citet{Jewitt13} detected a comet-like dust trail on active asteroid ((3200) Phaethon) and found 
the size of the dust particles  ($\sim 1\mu m$) is much smaller than that of the ordinary IPD  ($\sim 10\mu m$).

The dust cloud we detected could be formed by the collision of asteroids or 
the dust trail from the active asteroids (or main-belt comets). The dust particles in the newly formed dust cloud are 
exerted to the radiation pressure of the Sun. Sub micron particles are 
expelled from the solar system, while large particles suffer
the Poynting-Robertson drag \citep{Burns79} and spiral into the Sun. The lifetime is proportional to the size and the square of
the radius of the orbit. The typical life time of a spherical particle of $1\mu m$ radius at 3 au
is several thousand years. If the dust cloud we detected is formed very recently, there remains a large amount of small dust particles
in the cloud, as is observed for (3200) Phaethon \citep{Jewitt13}. These small particles could be
responsible for the blue color observed at $r=2.94$ au. 

In the main belt, similar dust clouds could exist and the distribution of the IPD could not be homogeneous.
A future deep-space probe with a state-of-the-art photometer, such as EXZIT  \citep{Matsuura14} will 
observe the distribution of dust clouds in the main belt, and delineate the origin of the IPD.

Recently, Kawara et al. (2017) examined the optical EBL using HST data, and found bright isotropic emission. They
attributed this emission to the isotropic zodiacal light, since their spectra are very similar. 
The brightness of this isotropic emission amounts to 80 $nWm^{-2}sr^{-1}$ in both the {\it {\it B}} and {\it R} bands. If
such isotropic emission exists surrounding the Earth, we may see its signature in Figures 4 and 5 as an abrupt
change or discontinuity in the observed brightness. On the other hand, since the present analysis uses only
differential brightness relative to that of the high ecliptic latitudes, the exact significance of such a signature in
Figures 4 and 5 depends strongly upon the spatial distribution of isotropic emission. Further discussion of this issue
requires models of isotropic emission to be constructed, which is beyond the scope of this paper.

We found that the zodiacal light observed with the IPP on board Pioneer 10 confirms the validity of  the model
\citep{Kelsall98} at $r < 2 $ au. This is a very important finding for the study of
the EBL. Many researchers (\citet{Cambrecy01}; \citet{Levenson07};
\citet{Matsumoto05}; \citet{Matsumoto15}; \citet{Sano15}; \citet{Sano16}; \citet{Tsumura13b}; \citet{Matsuura17} etc.)
have tried to detect the near-infrared EBL by applying the model zodiacal 
light and found the excess near-infrared EBL ($30\sim50$ $nWm^{-2}sr^{-1}$ at the {\it J} band)
that cannot be explained through known sources. The local dust cloud found in this analysis does not 
cause any effects on these results, because the ecliptic plane was not used in these analyses. 
A detection of excess emission has not been widely approved because they use the model to subtract
the bright foreground emission component; the zodiacal light. Furthermore, the similarity of the
spectrum of the excess EBL to that of the zodiacal light suggests that the zodiacal light
may not be fully subtracted due to the model uncertainty \citep{Dwek05}. Our result confirming the model
\citep{Kelsall98}  even outside the Earth orbit secures its validity
and makes the EBL detection more reliable.

\section{Conclusion}
We have reanalyzed the IPP data from Pioneer 10, and obtained the following findings.

\begin{enumerate}
\item The zodiacal light at $r < 2$  au is well consistent with the model zodiacal light \citep{Kelsall98}. The
reflectances in the {\it {\it B}} and {\it R} bands are marginally consistent with the previous work.
\item At $r=2.94$ au, the zodiacal light is bluer than that at $r < 2$  au and the spatial distribution of the dust cloud clearly 
different from the model.  This indicates that Pioneer 10 encountered a local dust cloud at $r=2.94$ au which could be a transient phenomena 
caused by the collision of asteroids or dust trail from the active asteroids.
\item Our analysis confirms the validity of the model \citep{Kelsall98} and justifies the detection of the excess near-infrared 
EBL. 
\end{enumerate}

\acknowledgments
\section*{ACKNOWLEDGMENTS}
We thank M. Ishiguro and T. Ootsubo for the valuable discussions. K.T. is supported by JSPS 
KAKENHI grant Nos. 26800112 and 17K18789.

\clearpage

\begin{figure}
\figurenum{1}
\epsscale{0.8}
\plotone{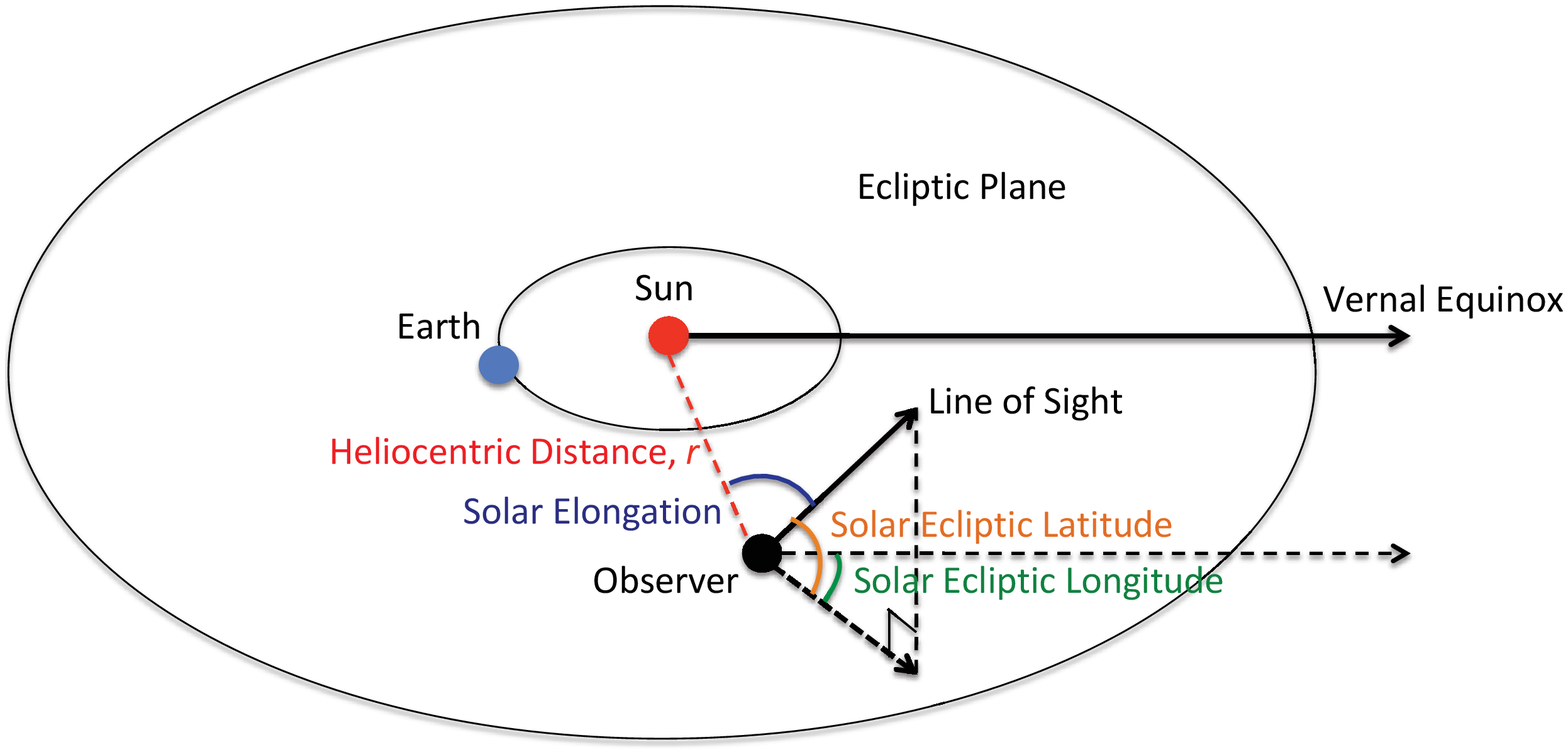}
\caption{Geometry  of the IPP observation and the coordinate system.      \label{fig1}} 
\end{figure}

\clearpage

\begin{figure}
\figurenum{2}
\epsscale{0.8}
\plotone{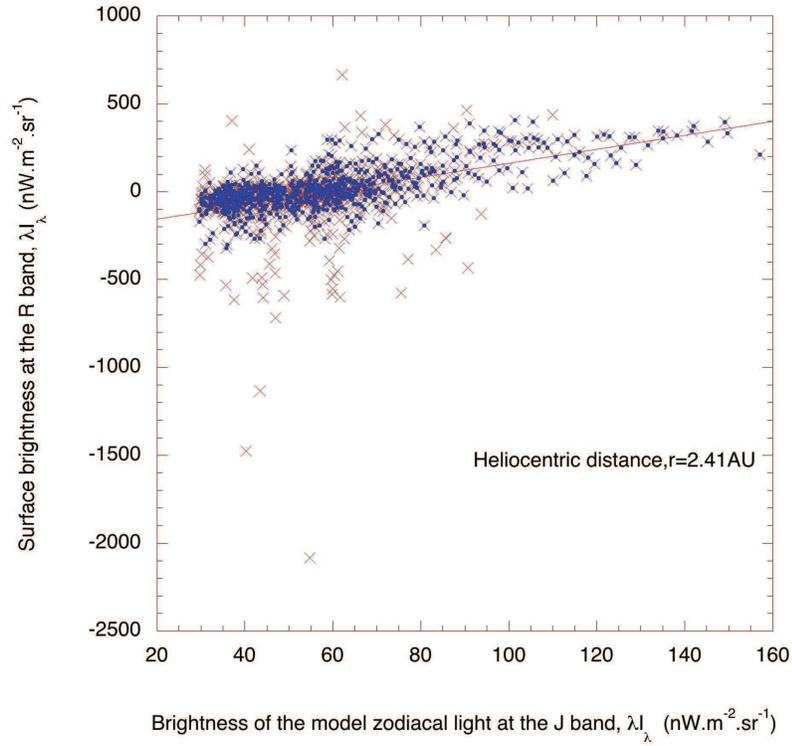}
\caption{Correlation diagram between the {\it R}-band residual brightness (red crosses) and the {\it J}-band model brightness  \citep{Kelsall98} at
 $r=2.41$ au. The straight line shows the result of the linear fit, and the filled blue circles show the dataset after clipping
 datapoints with brightnesses of three $\sigma$ over the fitted line.  \label{fig2}} 
\end{figure}

\clearpage

\begin{figure}
\figurenum{3}
\epsscale{0.8}
\plotone{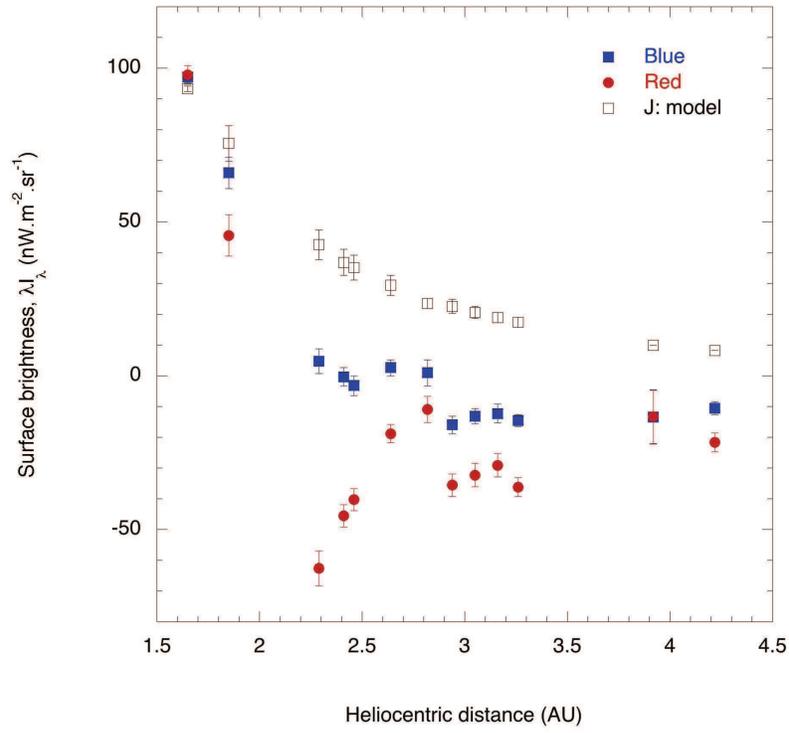}
\caption{ The average sky brightness at the ecliptic latitude, $|\beta|> 40^{\circ}$, and the solar elongation angle, $\varepsilon>90^{\circ}$ 
are plotted along the heliocentric distance, $r$. The blue squares, red circles, and open black squares indicate the results for the {\it B}-band, 
the {\it R}-band, and the {\it J}-band model brightness, respectively. The vertical bars of the {\it J} band are not the errors but indicate the scatter 
of the data point. \label{fig3}} 
\end{figure}

\clearpage

\begin{figure}
\figurenum{4}
\plotone{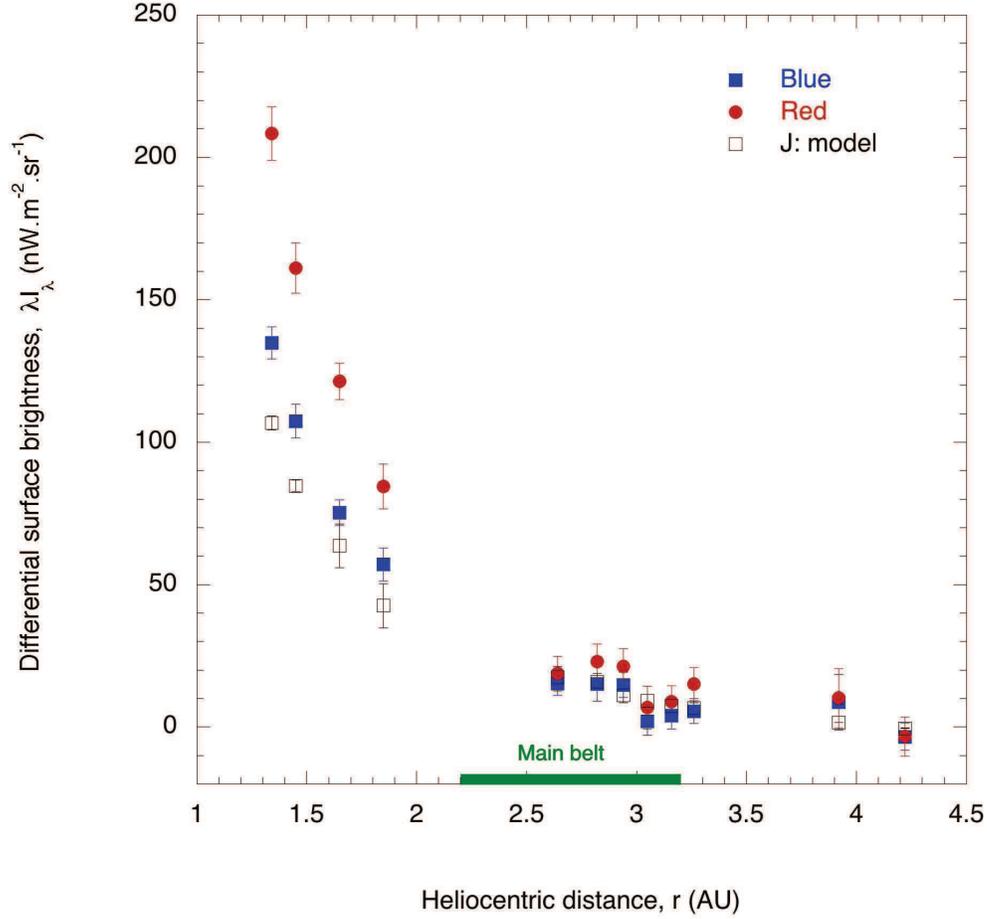}
\caption{ Dependence of the differential brightness of the zodiacal light upon the heliocentric distance. 
The average differential brightnesses on the ecliptic plane
($120^{\circ}<\varepsilon<150^{\circ}$, 
$|\beta|<10^{\circ}$) are plotted for the {\it B}-band (blue filled squares), the {\it R}-band (red filled squares), and
the {\it J}-band model brightness (black open squares). The thick green line on the horizontal axis shows the position 
of the main belt (the asteroid belt). \label{fig4}}   
\end{figure}

\clearpage

\begin{figure}
\figurenum{5}
\plotone{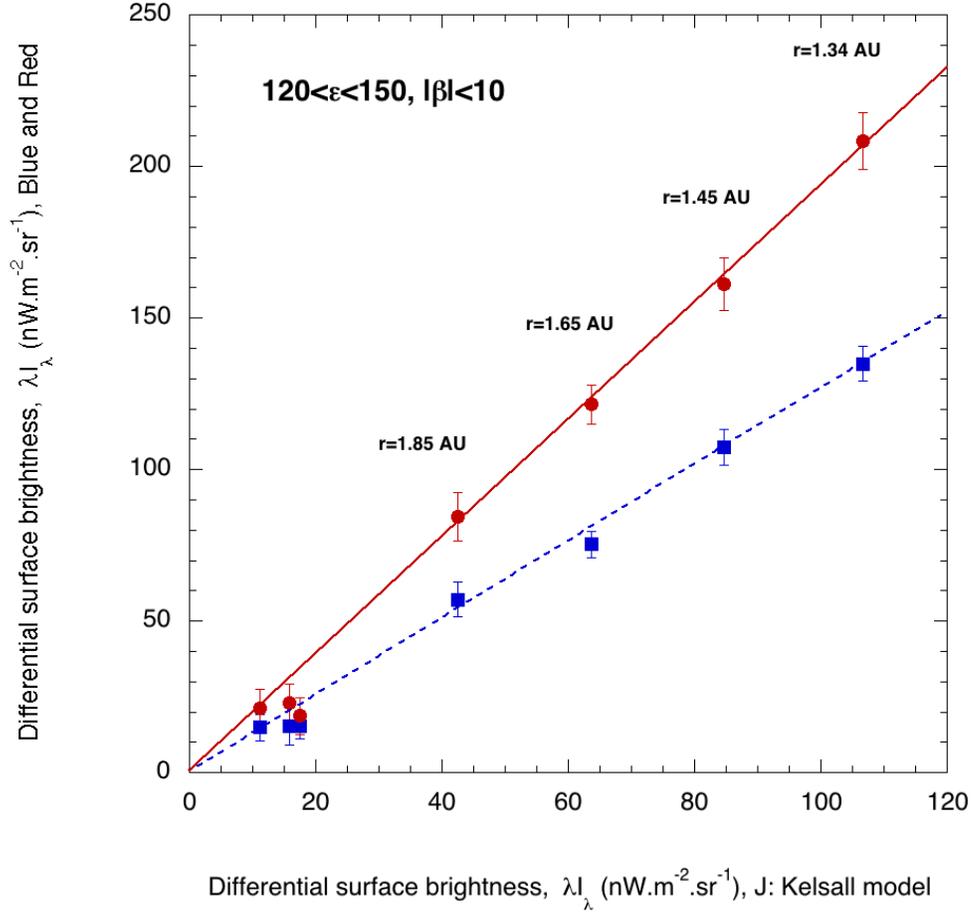}
\caption{Correlation diagram between the observed zodiacal light and the {\it J}-band model brightness. The blue filled squares and 
the red filled circles represent the {\it B}-and {\it R}-band  brightness, respectively. The red straight line and the blue dotted line 
indicate the linear fit results  for the {\it R}-and {\it B}-band brightnesses, respectively. \label{fig5}}
\end{figure}

\clearpage

\begin{figure}
\figurenum{6}
\plotone{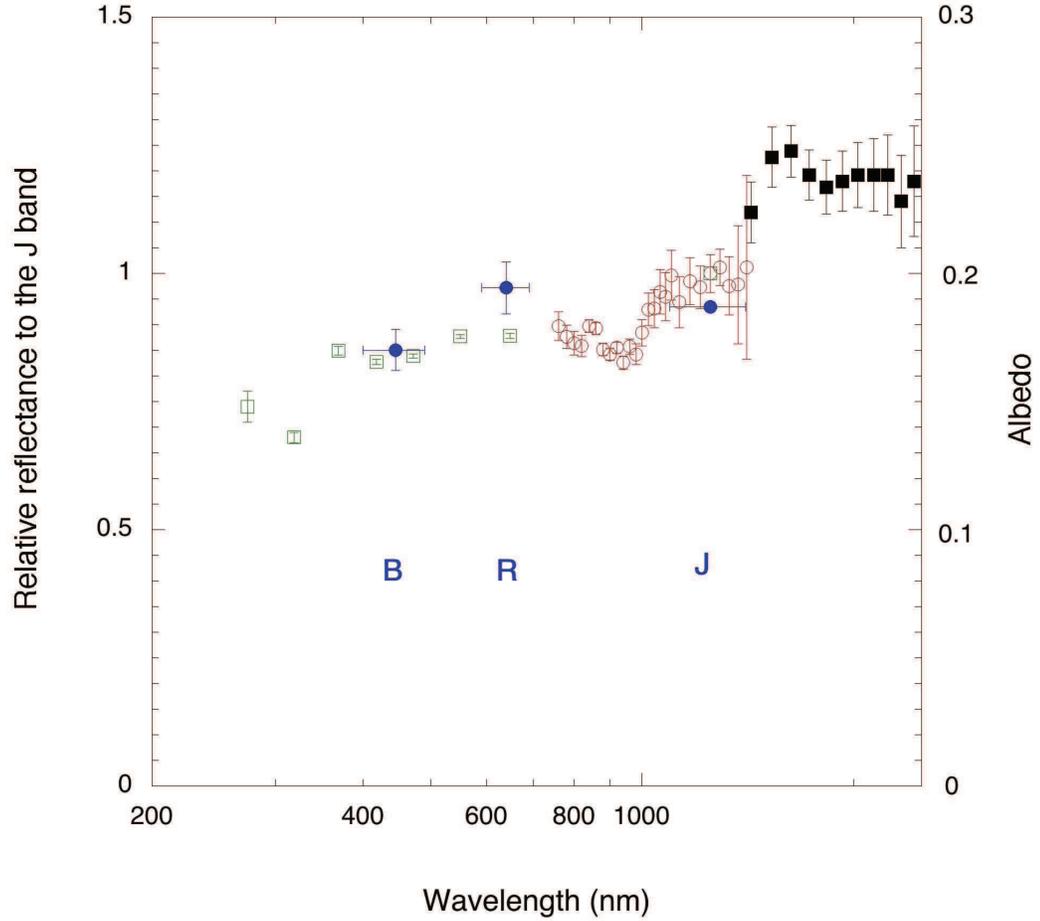}
\caption{Summary of the observed reflectances of the IPD. Unit of reflectance is arbitrarily normalized to be 1.0 at the {\it J} band 
for consistency with other works.
The albedo is shown in right ordinate based on the model \citep{Kelsall98} assuming that the scattering functions are same 
in this wavelength range.
The blue filled circles show results from the present work, while black filled squares, 
red open circles, and green open squares show the results by \citet{Matsumoto96}, \citet{Tsumura10}, and \citet{Kawara17}, respectively.
The horizontal bars of our result indicate the bandwidth.  
 \label{fig6}}
\end{figure}

\clearpage

\begin{figure}
\figurenum{7}
\plotone{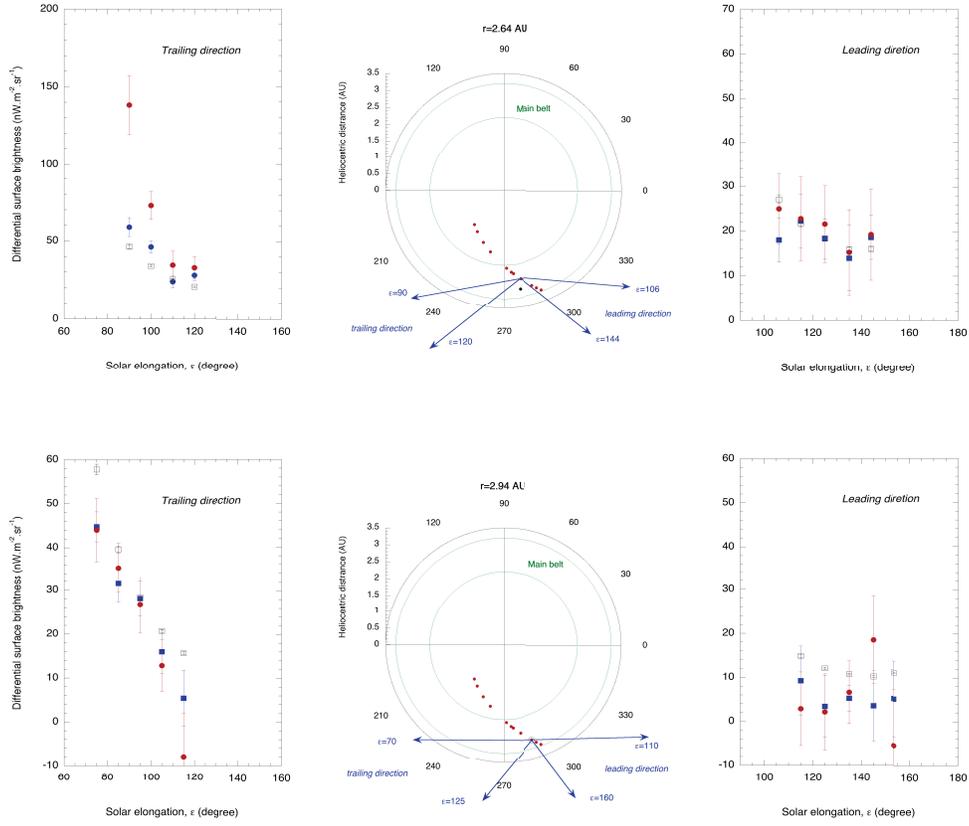}
\caption{Dependence of the zodiacal light upon the solar elongation angles at $r = 2.64$ (top) and $r = 2.94$ au (bottom). 
The average brightnesses for every 10 degrees at $|\beta|<10^{\circ}$ are plotted for the {\it B}-band (blue filled squares), 
the {\it R}-band (red filled squares), and the {\it J}-band model (black open squares). The right and left panels correspond to the leading and
trailing directions, respectively. The central parts shows the trajectory of Pioneer 10 and line of sights of the IPP at $r = 2.64$ au (top)
and $r = 2.94$ au (bottom). The filled black circle in top middle figure indicates the restored 
position of Pioneer 10 at $r=2.94$ au to the observation epoch at  $r=2.64$ au, assuming the Kepler motion.}  \label{fig7}
\end{figure}

\clearpage

\begin{figure}
\figurenum{8}
\plotone{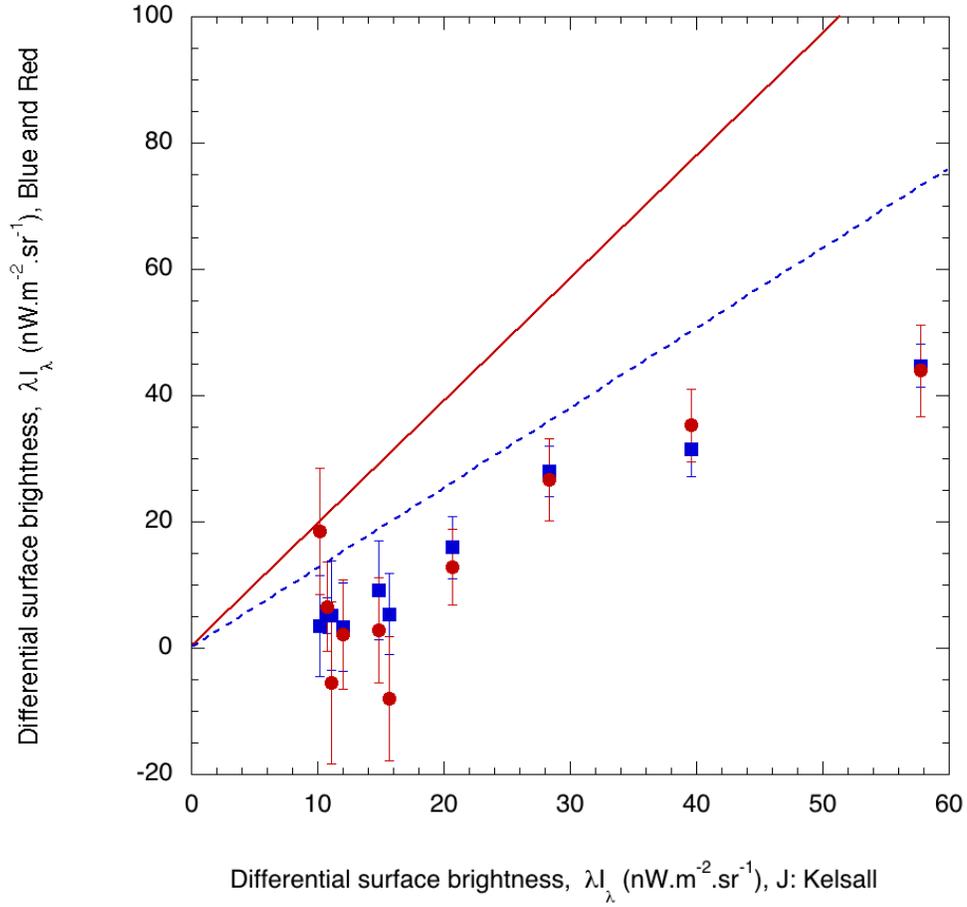}
\caption{Correlation diagram between the observed zodiacal light and the {\it J}-band model brightness at $r = 2.94$ au. 
The same data as Figure 7 are plotted. The blue 
filled squares and  red filled circles represent the {\it B}-and {\it R}-band  brightnesses, respectively. The red straight line and blue 
dotted line indicate the same ones in Figure 4. \label{fig8}}
\end{figure}

\clearpage

\end{document}